\documentclass[prl,twocolumn,aps]{revtex4}

\newcommand{\ba}{\begin{eqnarray}}
\newcommand{\ea}{\end{eqnarray}}
\newcommand{\be}{\begin{equation}}
\newcommand{\ee}{\end{equation}}

\newcommand{\ed}{\end{document}}
\newcommand{\nn}{\nonumber\\}

\begin{document}

\title{ Issues of duality on non-commutative manifolds: the {\it non-equivalence} between
self-dual and topologically massive models}

\author{T.Mariz$^{a}$, R. Menezes$^{a}$, J.R.S. Nascimento$^{a}$, R.F.Ribeiro$^{a}$ and C. Wotzasek$^{b}$}

\affiliation{$\mbox{}^{a}$Departamento de F\' \i sica, Universidade
Federal da Para\'\i ba, 58051-970 Jo\~ao Pessoa, Para\'\i ba, Brazil.
\\ $\mbox{}^{b}$Instituto de F\'\i sica, Universidade Federal do Rio de
Janeiro, 21945, Rio de Janeiro, Brazil}

\begin{abstract} 

We study issues of duality and dual equivalence in non-commutative manifolds.  In particular the question of dual equivalence for the actions of the non-commutative extensions of the self-dual model (NC-SD) in 3D space-time and the Maxwell-Chern-Simons model (MCS-SD) is investigate. We show that former model {\it is not} dual equivalent the non-commutative extension of the Maxwell-Chern-Simons model, as widely believed, but a to deformed version of it that is disclosed here.
Our results are not restrict to any finite order in the Seiberg-Witten expansion involving the non-commutative parameter $\theta$.

\end{abstract}

\maketitle

This paper is devoted to study the notions of duality transformations, self-duality and duality equivalence on non-commutative manifolds. This seems necessary because non-commutativity introduces new elements in the analysis that demands further investigations regarding already known phenomena and the eventual appearance of new physics.
Recent developments in string theory and mathematics has motivated the study of field theory models constructed on non-commutative spacetime.
Field theory models constructed on such spaces have many interesting features which their commutative counterparts do not share, like the possibility of novel
soliton solutions, UV/IR mixing and loss of the duality correspondence basically because of the presence of non-linear and non-local terms introduced by non-commutativity of the space-time.

The study of duality is done here using elements of the dual projection approach - a canonical transformation that separates the dynamical sector of a given theory from the sector that displays the symmetry of the action.
This approach has proved useful in disclosing duality mapping and model equivalence in the past \cite{embedding,EMCA}, in a different context.
In Ref.\cite{embedding} in particular, it has been shown that, in the ordinary commutative space, the Maxwell-Chern-Simons action (MCS) can be diagonalized in two sectors: one sector is described by the Self-Dual (SD) model that carries the self-dual dynamics known to exist in the MCS model but is not gauge invariant while the other sector is described by a pure Chern-Simons (CS) theory that, due to its topological character, has no propagating degree of freedom but carries the gauge symmetry of the original model. This procedure is here adapted to study duality in non-commutative manifolds.

Our approach is not based on the perturbative structure of the Seiberg-Witten map.  Consequently our final answer results to be non perturbative in the non-commutative constant $\theta$. This is distinct to other approaches up to now.
We clearly show that the expected and largely believed dual equivalence between a
non-commutative extension of the self-dual model (NC-SD) and the non-commutative version of the topologically massive theory (NC-MCS) is not realized in the conditions considered here. In order to support our findings we compute explicitly the gauge invariant, second-order model dual to the NC-SD model.

The investigation in non-commutative field theory
has experienced an outburst of interest after the work of Seiberg
and Witten \cite{SW} in open string and D--brane physics. A great
deal of effort is still under way in order to compute the relevant
modification and interpret the physical consequences that space non
commutativity effects have over known phenomena. Among the topics under
investigation, it becomes quite important to understand to which
degree duality is preserved or modified under the restriction
imposed by non-commutativity. Duality, as is well known, is a
symmetry concept that is very useful in different areas of physics
with dramatic consequences, quite particularly in gauge field
theories where it has been used to establish (i) dual equivalence
between solitonic solution of a given model with the particle
excitations of its dual and (ii) to disclose the presence of the
hidden gauge symmetry in self-dual solutions of some planar
models, as shown by Deser and Jackiw \cite{DJ}, with consequences
in the long distance physics of massive gauge theories. It is
therefore of fundamental importance, in order to extend the use of
these concepts to this new arena, to study the presence of duality
in non-commutative manifolds since it acts upon the theory in a
quite distinct manner.

The crucial point seems to be the
Seiberg-Witten map \cite{SW}, since duality has been approached exclusively
upon this concept \cite{ganor}. The rationale behind this
technique is the mapping of the non-commutative gauge fields into
ordinary commutative ones in powers of the non-commutative
parameter $\theta$ in this way establishing a connection between the
gauge orbits in both regimes and has been
extended to tensors of distinct nature \cite{VR}.  This idea, originally applied in
the study of duality in {NC} electromagnetic phenomena \cite{ganor} was extended to establish the duality equivalence between different
mathematical descriptions of some known planar phenomena in
ordinary 3D space-time \cite{many}.

It is shown in the present report, via the dual projection technique, that a simple and beautiful solution of the duality problem that avoids the use of the Seiberg-Witten expansion is possible.
Besides the non-perturbative character of the solution, this approach which is derived from a Noether embedding algorithm, makes
no use of a gauge invariant master action or the Lagrange multiplier imposing zero-curvature condition approach to establish duality.

Let us consider the Maxwell-Chern-Simons action, defined over a non-commutative manifold (for a review of the non-commutative field theory, see \cite{DouglasNekrasov,Szabo})
\ba &&{\cal S}[A] \!\!= \!\!\int\mathrm{d}^{3}x \!\!\left[-\frac{1}{2}{ F}_{\mu} \star{ F}^{\mu} \right.\nn
&+&\!\!\!\!\left.
\frac{m\chi}{2}\epsilon^{\mu\nu\lambda}\!\!\left(\!\!{A}_{\mu}\!\star\!
\partial_\nu {A}_{\lambda}\!\! +\!\! \frac{2i}{3}{ A}_{\mu}\!\star\!{
A}_{\nu}\!\star\!{ A}_{\lambda} \!\!\right)\!\!\right] \label{380}
\ea
where $m$ is a free parameter with mass dimensions. We used that $F_\mu = \frac 12 \epsilon_{\mu\nu\lambda} F^{\nu\lambda}$ and $F^{\nu\lambda} = \partial^{[\nu} A^{\lambda ]} + i [A_\nu , A_\lambda]_{\star}$. The $\star$-product is given by the Moyal formula,
\be
 ( f \star  g)(x) =
 \left.
 e^{\frac i 2{\theta_{\mu\nu}}\partial_{\xi_\mu}
 \partial_{\zeta_\nu}} f(x+\xi)  g(x+\zeta)
 \right\vert_{\xi=\zeta=0}
 \label{moyal}
\ee
and the non-commutativity of the space-time is described by,
\begin{equation}
[x^{\rho},x^{\sigma}]_{\star}=i\theta^{\rho\sigma} .
\label{nc}
\end{equation}
The physics behind the presence of the non-commutative Chern-Simons term has been extensively studied in recent years \cite{Krajewski,BichlGrimstrup,ChenWu,Polychronakos,BakKim,Sheikh1,BakLee,NairPolychronakos,Chu,GrandiSilva}.

Our goal is, as discussed in the introduction above, to look for
the NC dual of the theory (\ref{380}).
To apply the dual projection algorithm into the above model one needs first to include an ancillary field, say $\pi_\mu$, to duplicate the dimension of the phase space or, equivalently, lower the order of the differential equations. The Lagrangian density for NC-MCS model becomes
\ba {\cal L}[A]\!\!\!&=& \!\!\! \left[\frac{m^2}{8}(\pi_\mu - A_\mu)^2_{\star}+ \frac{m\chi}{2}\epsilon^{\mu\nu\lambda}{\pi}_{\mu}\!\star\!
\partial_\nu {A}_{\lambda} \right.\nn
&+&\left.  \frac{m\chi\, i}{6}\epsilon^{\mu\nu\lambda}(3 \pi_{\mu}- A_\mu)\!\star\!{
A}_{\nu}\!\star\!{ A}_{\lambda}\right] \label{380a}
\ea
As shown in \cite{embedding}, the final effect of the dual projection algorithm is materialized by the following canonical transformation that transforms the phase space of this model back into the configuration space of two-field theory, albeit of first-order,
\be
\label{canonical transformation}
\pi_\mu = f_\mu^+ + f_\mu^- \;\;\; ; \;\;\; A_\mu = f_\mu^+ - f_\mu^-
\ee
Plugging this field redefinition back into NC-MCS action, we obtain
\ba
&&{\cal L}[A]\!\! = \!\!\frac m2\! \left[\,\epsilon^{\mu\nu\lambda}\left(f^+_\mu\!\star\!
\partial_\nu f^+_\lambda\!\! +\!\! \frac{2i}{3}f^+_\mu\!\star\! f^+_\nu\!\star\!f^+_\lambda\right)\right. \nn
&& \!\!+ m \!\left(f_\mu^-\right)^2_\star -\!\epsilon^{\mu\nu\lambda}\!\!\left(\! f^-_\mu\!\star\!
\partial_\nu f^-_\lambda\!\! -\!\! \frac{4i}{3}f^-_\mu\!\star\! f^-_\nu\!\star\! f^-_\lambda\!\right)\nn
&& - \left. i\, 2 \,\epsilon^{\mu\nu\lambda}\left(f^-_\mu
\star f_\nu^-\star f_\lambda^+\right) \right]\, .
\label{dual projection}
\ea
Therefore, following the steps of ref.\cite{embedding} we have rewritten the NC-MCS in terms of two components fields, $f_\mu^{\pm}$, presenting the desired features: a pure Chern-Simons term for the $f_\mu^+$, and a self-dual-like model composed of a mass term and a Chern-Simons term-like term. There are however important issues to overcome. One is that the cubic factor in the CS-like term has the wrong coefficient. The second and most important difficulty is the presence of a mixed cubic term (the last term in (\ref{dual projection})).  It shows that the original NC-MCS cannot be diagonalized in a dynamical sector plus a topological sector displaying the original symmetry which seems to be a hard nut to crack. Therefore, differently from the commutative case, here the NC-MCS theory {\it is not} dual to the NC-SD theory.  This is the first part of our result. Next we try to fix the above mentioned problems.

It is important to observe at this juncture that a deformation of the Chern-Simons term is of no help to solve any of the above difficulties. A close inspection in expression (\ref{380a}) shows that the cubic terms originate from two contributions: an $A^3_{\star}$  and a $\pi\star A^2_\star$ terms. A more symmetric combination seems to be more adequate and indeed, a little theoretical experimentation shows that the inclusion of another cubic term in the form of $i\,\alpha A\star\pi^2_\star $, with $\alpha$ being an adjustable constant tensor, is able to cure both sickness at once.  It should be observed that the presence of such a term demands a profound modification on the structure of the theory  since the substitution
$\pi_\mu\star\pi^\mu \to \pi_\mu\star(1 + i\,\alpha A)^{\mu\nu}\star\pi_\nu$ requires an initial structure in the second-order Lagrangian density as
${ F}_{\mu}\star{ F}^{\mu} \to{F}^{\mu} \star(1 + i\,\alpha A)^{-1}_{\mu\nu}\star{ F}^{\nu}$. The deformation therefore should be in the Maxwell part.

Let us then consider the following Lagrangian density for a {\it deformed} NC-MCS theory
\ba &&{\cal L}[A] \!= \! \left[-\frac{1}{2}{ F}^{\mu} \star\left(\mbox{}^* M_{\mu\nu}\right)\star{ F}^{\nu} \right.\nn
&+&\!\!\!\!\left.
\frac{m}{2}\epsilon^{\mu\nu\lambda}\!\!\left(\!\!{A}_{\mu}\!\star\!
\partial_\nu {A}_{\lambda}\!\! +\!\! \frac{2i}{3}{ A}_{\mu}\!\star\!{
A}_{\nu}\!\star\!{ A}_{\lambda} \!\!\right)\!\!\right] \label{380b}
\ea
The first-order Lagrangian density for the deformed NC-MCS model becomes then
\ba
&&{\cal L}[A]\!\!= \!\! \left[\frac{m^2}{8}(\pi \! - \! A)^\mu{\star}\left(\mbox{}^* M_{\mu\nu}\right)^{-1}\star(\pi\! -\! A)^\nu +  \right.\nn
&&\left. \frac{m}{2}\epsilon^{\mu\nu\lambda}\left({\pi}_{\mu}\!\star\!
\partial_\nu {A}_{\lambda} + \frac{i}{3}(3 \pi\! -\! A)_\mu\!\star\!{
A}_{\nu}\!\star\!{ A}_{\lambda}\right)\right]\nonumber \label{380c}
\ea
where
\be
\left(\mbox{}^* M_{\mu\nu}\right)^{-1}=\delta_{\mu\nu} - \frac{2\, i}{m}\epsilon_{\mu\nu\lambda}A^\lambda
\ee
Following the previous steps and plugging the field redefinition (\ref{canonical transformation}) back into the deformed NC-MCS action, we obtain
\ba
&&{\cal L}[A]\!\! = \!\!\frac m2\! \left[\,\epsilon^{\mu\nu\lambda}\left(f^+_\mu\!\star\!
\partial_\nu f^+_\lambda\!\! +\!\! \frac{2i}{3}f^+_\mu\!\star\! f^+_\nu\!\star\!f^+_\lambda\right)\right. \nn
&& \left.\!\!+ m \!\left(f_\mu^-\right)^2_\star \!-\!\epsilon^{\mu\nu\lambda}\!\!\left(\! f^-_\mu\!\star\!
\partial_\nu f^-_\lambda\!\! +\!\! \frac{2i}{3}f^-_\mu\!\star\! f^-_\nu\!\star\! f^-_\lambda\!\right)\! \right]\nonumber
\label{dual projection2}
\ea
which is our main result.  The first line displays a pure NC-CS term while the second line shows a NC-SD model displaying the same helicity of the original NC-MCS theory.  Clearly the original gauge symmetry is supported by the first sector while dynamics is manifest by the second as discussed above.  Therefore we were able to obtain the thoroughly searched dual of the NC-SD model as manifested by the {\it deformed} NC-MCS, not the original theory. It is important to mention that this is an exact result.  

This result is a very important contribution to the present stage of investigation on this subject since it shines light on this debated matter by bringing in an exact result for an interesting model and a new procedure.  It is also of importance to the bosonization program in the NC space.  As shown in \cite{ghosh1} it is possible to establish a connection between the massive Thirring model and the NC-SD model, at least in the long wave length limit.  Our results clearly show that, contrary to what would be naively expected, there is no connection from this model to the NC-MCS model.  We showed that NC-MTM is, in fact, dual to the model proposed in (\ref{380b}).
Besides, since non-abelian gauge theories have a similar gauge structure as the non-commutative gauge theories, the study of duality and bosonisation in non-commutative space-times is also of interest as the studies of the latter may shed further light to the similar problems in the former.

In summary, we have computed directly the dual equivalent of the NC-SD model.  Our results show that the NC-SD model {\it is not} dual to the NC-MCS model as thoroughly searched but to a {\it deformed} version of it as presented in Eq.(\ref{380b}).  The diagonalization of this action was possible and displays the presence of a pure NC-CS term carrying the gauge symmetry plus a dynamical sector in the form of a NC-SD model. Our result clearly explains the reasons for the failure in obtaining duality equivalence through the master action approach and other methods, being limited therefore to the first-order expansion in the Seiberg-Witten map. 
Although we have found it in a very particular model it seems to be a quite general phenomenon in NC manifolds.  This is a new result that illuminates the discussion on the subject and may have further consequences on the studies of duality on NC manifolds.

\acknowledgments

We would like  to thank CAPES, CNPq, PROCAD and PRONEX for financial support.

\end{document}